\pdfoutput=1
\documentclass[sigconf,screen]{acmart}

\usepackage{amsmath,newtxtext,newtxmath}
\usepackage{colortbl}
\usepackage{algpseudocode}
\usepackage{algorithm}
\usepackage{graphicx}
\usepackage{textcomp}

\usepackage{caption}
\usepackage{subcaption}
\usepackage{soul}
\usepackage[inline,shortlabels]{enumitem}
\usepackage[capitalise]{cleveref}
\algrenewcommand{\algorithmicrequire}{\textbf{Input:}}
\algrenewcommand{\algorithmicensure}{\textbf{Output:}}
\usepackage{algorithm}
\usepackage{url}
\usepackage{float}
\usepackage{afterpage}
\usepackage{breakurl}
\usepackage{dblfloatfix}
\usepackage{balance}

\copyrightyear{2023}
\acmYear{2023}
\setcopyright{rightsretained}
\acmConference[SEC '23]{The Eighth ACM/IEEE Symposium on Edge Computing}{December 6--9, 2023}{Wilmington, DE, USA}
\acmBooktitle{The Eighth ACM/IEEE Symposium on Edge Computing (SEC '23), December 6--9, 2023, Wilmington, DE, USA}
\acmDOI{10.1145/3583740.3626611}
\acmISBN{979-8-4007-0123-8/23/12}

\begin{document}

\title{CWASI: A WebAssembly Runtime Shim for Inter-function Communication in the Serverless Edge-Cloud Continuum}


\author{Cynthia Marcelino}
\email{c.marcelino@dsg.tuwien.ac.at}
\orcid{0000-0003-1707-3014}
\affiliation{%
  \institution{Distributed Systems Group, TU Wien}
  \city{Vienna}
  \country{Austria}
}

\author{Stefan Nastic}
\orcid{0000-0003-0410-6315}
\email{snastic@dsg.tuwien.ac.at}
\affiliation{%
  \institution{Distributed Systems Group, TU Wien}
  \city{Vienna}
  \country{Austria}
}

\renewcommand{\shortauthors}{Marcelino and Nastic}

\begin{abstract}
 Serverless Computing brings advantages to the Edge-Cloud continuum, like simplified programming and infrastructure management. In composed workflows, where serverless functions need to exchange data constantly, serverless platforms rely on remote services such as object storage and key-value stores as a common approach to exchange data. In WebAssembly, functions leverage WebAssembly System Interface to connect to the network and exchange data via remote services. As a consequence, co-located serverless functions need remote services to exchange data, increasing latency and adding network overhead. 
To mitigate this problem, in this paper, we introduce CWASI: a WebAssembly OCI-compliant runtime shim that determines the best inter-function data exchange approach based on the serverless function locality. CWASI introduces a three-mode communication model for the Serverless Edge-Cloud continuum.
This communication model enables CWASI Shim to optimize inter-function communication for co-located functions by leveraging the function host mechanisms. 
Experimental results show that CWASI reduces the communication latency between  the co-located serverless functions by up to 95\% and increases the communication throughput by up to 30x.
\end{abstract}

\ccsdesc[500]{Software and its engineering~Abstraction, modeling and modularity}
\ccsdesc[500]{Software and its engineering~Message passing}
\ccsdesc[500]{Computer systems organization~Cloud computing}

\keywords{WebAssemly, Inter-function, Edge-Cloud Continuum, Serverless, Shim}


\maketitle

\section{Introduction}
\label{sec1}

The Edge-Cloud continuum leverages many benefits from the Serverless Computing paradigm. Serverless Computing simplifies application development by removing the complexities of infrastructure management from the developer~\cite{baresi, one-step, Challenges-and-Opportunities}. Recently, the integration of WebAssemly (Wasm) with Serverless Computing enabled enhanced portability, improved performance, and broader language support in deploying serverless functions. Furthermore, in an environment composed of resource-constrained devices such as the Edge-Cloud continuum, Wasm offers many advantages due to its near-native execution speed, security, and reduced cold start. In Wasm, large container images are replaced with small binary files executed in isolated sandboxes. Wasm creates a secure sandbox typically without host access, increasing security and privacy, which leads to additional protection for co-located guest applications~\cite{Challenges-and-Opportunities,bringing-speed-wasm, Pushing}. Despite these advantages, Wasm's constrained host access introduces additional challenges to the complex serverless inter-function communication. 


\textbf{Data exchange between serverless functions.} 
The increased usage of serverless functions compositions exposed limitations regarding data exchange between serverless functions~\cite{one-step,scf,ristov_xafcl}. Currently, there are two main approaches enabling data exchange between serverless functions:
\begin{enumerate} [leftmargin=0mm,wide=10pt,label=(\alph*)]
    \item \textit{Remote third-party services:} Most of the current serverless platforms are function-locality agnostic. They leverage remote services to exchange data between the functions providing flexibility and scalability. Such remote services used for serverless data exchange can be classified as follows:
    \begin{enumerate*} [(i)]
        \item \textit{In-memory} solutions such as key-value stores (KVS)~\cite{anna,cloudburstSF} or remote far memory solution~\cite{Jiffy} provide low latency. Nevertheless, they still rely on third-party services, creating data, memory, and network overhead, making them inadequate for the Edge-Cloud continuum, where resources can be limited.   
        \item \textit{Storage-based} solutions are most common approach to exchange data between functions~\cite{why-when-how}. However, they increase latency by pushing functions to repeatedly download and upload the data. Additionally, they increase costs with remote storage while limiting usage to a single Cloud provider~\cite{Scalia, Wang2020AnAD}.
    \end{enumerate*}
    In Wasm, \textit{remote services} usage became possible after the release of WASI, which introduced POSIX-like calls such as network access~\cite{wasi}. This enables the Wasm binaries to connect to remote services via the host network interfaces.
    Although remote services enable data exchange between serverless functions, for co-located functions they add unnecessary network traffic, create data and resource overhead, duplicate data serialization on the source and target, and consequently increase latency. Unfortunately, with these approaches, co-located functions cannot benefit from the function proximity since the data exchange is done via the remote services~\cite{floki}. Furthermore, remote services lead to additional costs in the typical ``pay-per-use'' model.
  
  \item \textit{Inter-function communication:} Other approaches enable serverless functions to leverage direct data exchange to decrease latency and avoid extra remote services~\cite{scf, Serverless-Edge-Computing}. Two common approaches for inter-function communication are:
    \begin{enumerate*} [(i)]
        \item \textit{Message Queues} such as SAND~\cite{sand} leverage publish/subscribe paradigm for inter-function communication.  In Wasm, WASI libraries provide network communication enabling message queue communication~\cite{wasi}. Even though message queues enable point-to-point inter-function communication, they still rely on external message brokers. This creates network and resource overhead and dependency on external solutions.
        
        \item \textit{Shared memory} approaches~\cite{process-as-a-Service, freeflow} exploit sharing of the same memory region to enable low-latency data exchange between the serverless functions. For coordination, such approaches use pipes~\cite{nightcore}, threads~\cite{faastlane}, or custom software isolation~\cite{faasm,faabric}.
        Although shared memory provides low latency communication, resource sharing between two different functions reduces the isolation offered by containerization~\cite{pipeDevice}. Additionally, all the functions must be started simultaneously, as the shared memory address space must be allocated before the processes start, increasing the memory footprint.
        In Wasm, \textit{shared-memory}-based communication is typically enabled by statically linking the modules at Wasm function startup. The static link creates a shared memory region in the Wasm Virtual Machine (VM) which makes it accessible between the modules. Once host runtimes explicitly link the modules, Wasm modules can also be reused~\cite{static_link}. Nevertheless, static linking is still unavailable in the current wasm-based Edge-Cloud infrastructure as it requires the modules to be explicitly statically linked via the host function. Current state-of-the-art wasm-based shims enable single Wasm module execution at a time.
    \end{enumerate*}
\end{enumerate}

\textbf{Contributions}. In this paper, we introduce CWASI, a runtime shim that facilitates inter-function communication between wasm-based serverless functions. CWASI is a container runtime shim that leverages Wasm runtimes to provide isolation and security. CWASI optimizes inter-function communication for co-located serverless functions by introducing mechanisms that select the best inter-function communication approach.
The main contributions of this paper include: 
\begin{itemize}
    \item \textit{A novel model for serverless inter-function communication}
     that leverages function locality to enable three-mode communication. This model enables co-located functions to leverage the local host mechanisms to optimize inter-function communication, consequently reducing their dependency on external services and network connections;
    \item \textit{A WebAssembly Container Runtime Shim} that enables the three communication modes proposed by our inter-function communication model to optimize data exchange between serverless functions. By following the three-mode communication model, CWASI presents a reduction of up to 95\% in latency and an increase in throughput by up to 30 times.
\end{itemize}

\textbf{Outline}. 
The paper has eight sections. \cref{sec2} describes the motivating scenario and our research challenges.
\cref{sec3} discusses related work and limitations. 
\cref{sec4} gives an overview of the CWASI Communication Model and its Shim Architecture. 
\cref{sec5} describes the mechanisms introduced in CWASI and their usage. 
\cref{sec6} shows the implementation details.
\cref{sec7} describes the experiments and evaluation, and \cref{sec8} concludes with final discussion and future work.
\section{Motivation} 
\label{sec2}
\subsection{Illustrative Scenario}\label{mot:scenario}

To better motivate the research challenges, we present a realistic real-time video analytic use case for vehicle path reconstruction. A serverless workflow for vehicle path reconstruction typically involves the deployment of cameras at intersections and along the road. The cameras detect the vehicle by using object detection algorithms and then, serverless functions are employed to identify the vehicle's location in its path.
Our workflow features four serverless functions partially executed on the Edge and partially executed on the Cloud. The use case is based on AWS use cases~\cite{autonomous-driving} and scientific researches~\cite{react, Serverless-Video-Analytics, Bridging-the-edge-cloud}. To decrease the communication latency, large real-time video streams are processed on edge nodes responsible for extracting image frames and other tasks such as labeling and anonymization. In contrast, tasks that require greater computing resources, such as more powerful object detection models, are performed in the Cloud.

\begin{figure}[!htb]
\centering
\includegraphics[width=0.9\linewidth]{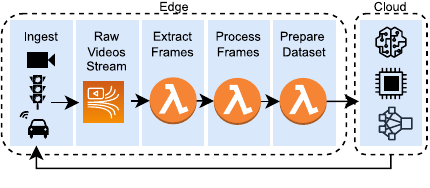}
\captionsetup{justification=centering}
\caption{Serverless Workflow for Image Processing for \\ Vehicle Path Reconstruction (simplified)}
\label{fig:batch_processing}
\end{figure}

In \cref{fig:batch_processing}, the \textit{Ingest} stage represents the real-time videos that the cameras collect and transmit to edge nodes via a streaming framework. \textit{Raw Videos Stream} stage triggers the serverless functions, which are responsible for \textit{Extract Frames} from the real-time videos. To decrease the latency, the real-time videos are processed in small chunks where each \textit{Extract Frames} function instance is responsible for processing a small part of the video. 
Once \textit{Process Frames} functions have finished labeling and anonymization, they again place the results from \textit{Process Frames} in the storage. In the next stage, the function \textit{Prepare Dataset} retrieves the images from the storage and places the results in the storage again. Further, the workflow continues in the cloud, where the function retrieves the prepared dataset results from \textit{Prepare Dataset} for more resource-intense tasks such as model training for object detection. At the end of our workflow, the trained model is returned to the edge nodes, which can act upon the newly trained model, such as recreating a vehicle path.

CWASI shim detects when \textit{Extract Frames}, \textit{Process Frames} and \textit{Prepare Dataset} functions are running on the same host. Thus, CWASI skips remote storage and applies co-located function communication optimization to exchange data for functions running on the same host. CWASI simplifies the workflow, decreases latency, and reduces financial costs with remote storage services.



\subsection{Research Challenges}
Current serverless platforms have limited support for inter-function communication. 
In wasm-based serverless approaches, the challenges are even more significant for co-located inter-function communication, given that WASI only allows limited access to the host. We identify the following main research challenges:

\textbf{RC-1: How can we enable efficient inter-function communication in wasm-based serverless platforms?} Inter-function communication is crucial for real-time scenarios where edge devices must quickly react upon scenario changes. In our scenario, inter-function communication latency directly affects the reaction time of the path reconstruction which can influence the vehicles' trajectory. In the given scenario, efficiency and latency are critical factors for a reliable workflow. However, current state-of-the-art serverless platforms have remote services as a common approach for exchanging data. This is a burden in the Edge-Cloud environment for composed workflows, which constantly exchange data. Serverless functions must repeatedly download and upload data, increasing resource footprint, network, and data access overhead which leads to a complexity increase of inter-function communication~\cite{one-step, Challenges-and-Opportunities, sledge}.

\textbf{RC-2: Can we utilize Wasm static-linking to optimize inter-function communication?} Wasm static-linking enables multiple modules to access the same memory region. Each module has a dedicated linear memory space. However, the modules can access other memory spaces within the same Wasm VM when statically linked, enabling shared memory and Wasm module reuse. To statically link the modules, every Wasm binary must be explicitly added in the Wasm VM by the host runtime before starting it. WebAssembly static linking is still unavailable in state-of-the-art Wasm shims such as RunWasi WasmEdge and Wasmtime~\cite{runwasi, static_link}. The lack of Wasm static linking exposes a challenge in achieving optimal communication between modules in the same serverless workflow.

\textbf{RC-3: Can we exploit function locality to optimize the communication for co-located Wasm serverless functions?} Co-located Serverless functions such as \textit{Extract Frames} and \textit{Process Frames} can profit from the function proximity to decrease communication latency. Function locality awareness enables us to leverage the host mechanisms, avoiding common challenges introduced by network inter-function communication, such as data duplication and redundant serialization on the source and target functions, besides resource and network overhead ~\cite{faasm, sledge, Challenges-and-Opportunities}.
\section{Related Work} \label{sec3}
\subsection{Serverless Platforms Models and Shims}

\begin{figure*}[!htb]
\centering
    \begin{subfigure}[t]{0.24\textwidth}
    \centering
    
    \includegraphics[width=\textwidth]{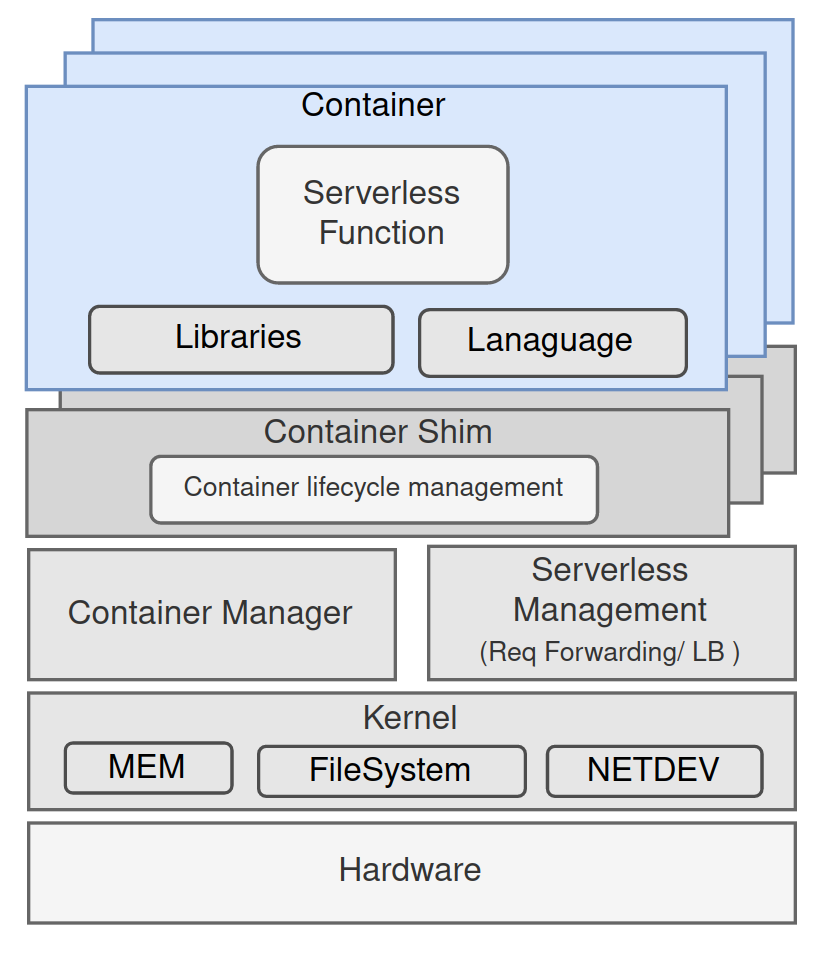}
    \caption{Container-based Serverless} \label{fig1_a}
\end{subfigure}\hfill
    \begin{subfigure}[t]{0.24\textwidth}
    \centering
    \includegraphics[width=\textwidth]{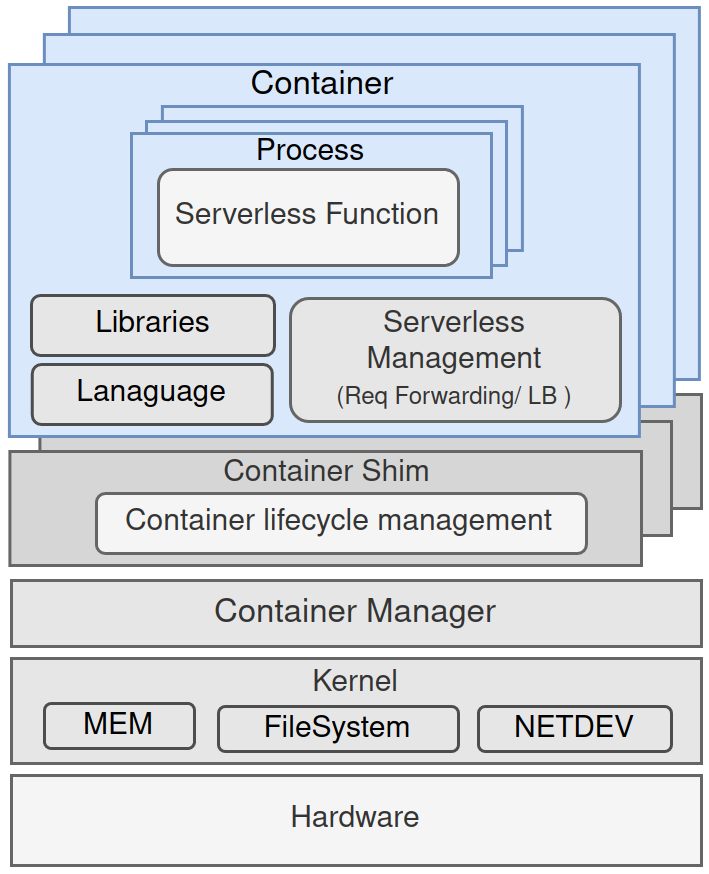}
    \caption{Process-based Serverless} \label{fig1_b}
\end{subfigure}\hfill
\begin{subfigure}[t]{0.24\textwidth}
    \centering
    \includegraphics[width=\textwidth]{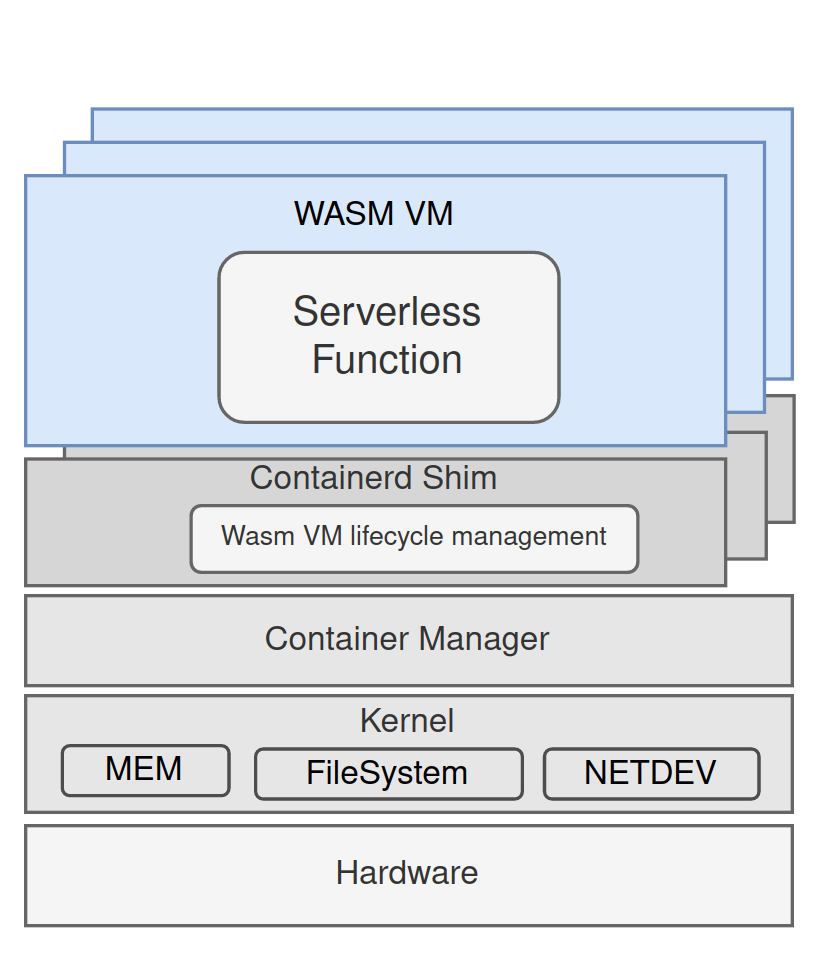}
    \caption{WebAssembly in the Cloud} \label{fig1_c}
\end{subfigure}\hfill
 \begin{subfigure}[t]{0.24\textwidth}
    \centering
    \includegraphics[width=\textwidth]{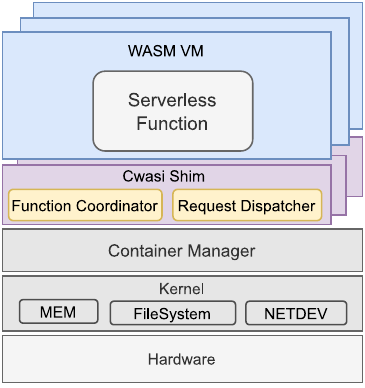}
    \caption{CWASI} \label{fig1_d}
\end{subfigure}\hfill

\caption{Overview of Edge-Cloud Serverless Platforms}
\label{fig:arch_overview}
\end{figure*}

CWASI is a WebAssembly runtime shim that interacts with container managers such as \textit{containerd} to manage containers' process lifecycle. \cref{fig:arch_overview} shows how CWASI fits into the existing Edge-Cloud stack. \cref{fig1_a} shows the standard container-based serverless platforms such as OpenWhisk, where every function is packed in a container, and a serverless management framework such as request forwarding is shared among all containers~\cite{Serverless-Computing-Design, Evaluation-of-OpenWhisk}. Container-based serverless offer strong isolation as every serverless function is packaged on its container with dedicated libraries and language runtime. Nevertheless, large container images with a single serverless function lead to a memory footprint overhead and high startup latency. Hence, it is inefficient for the Edge where resources are limited and low latency required~\cite{Challenges-and-Opportunities,faastlane, one-step, kumar2020performance}.

In process-based serverless stack, in \cref{fig1_b}, serverless frameworks such as OpenFaas introduces additional process in the standard container. Serverless functions share container resources such as libraries and language runtime with side-car serverless management processes~\cite{Understanding-Serverless, randazzo2019kata, Serverless-Edge-Computing}. For example, in OpenFaas, every function has an additional watchdog process acting as a reverse proxy~\cite{sledge, openfaas}. The serverless function and the watchdog are isolated processes but share the same container resources. Although process-based serverless platforms decrease the overhead compared to container-based platforms (\cref{fig1_a}), they still rely on containerization, which leads to resource overhead such as memory and cold start issues~\cite{Challenges-and-Opportunities, faasm, one-step}.

\cref{fig1_c} shows how container managers and runtime shim enables Wasm in the current Edge-Cloud infrastructure. Two key features contribute to adopting Wasm outside the web browser: WASI and Host Runtime~\cite{webAssembly-in-non-web, wasi, Wasmtime, TruffleWasm}. WASI provides access to the host machine and, consequently, to the network. Host runtime creates the Wasm VM by statically loading and executing the Wasm binary file enabling Wasm module interaction via imports and exports~\cite{wasi, Weakening-WebAssembly}. Wasm enables near-native speed and decreases serverless cold start issues significantly~\cite{Evaluating-WebAssembly, lightweight-design}. As Wasm binaries run on a specific sandbox with restricted host access, it presents limitations for inter-function communication.

We present CWASI approach in \cref{fig1_d}. A serverless runtime shim that leverages Wasm to provide isolation and security while enabling optimized inter-function communication. CWASI follows Open Container Initiative (OCI) specification to integrate Wasm in the Edge-Cloud continuum ~\cite{oci-spec}. Furthermore, CWASI introduces a novel approach to identify and select the best inter-function communication approach available for every serverless function based on the functions' location.

\subsection{Sandboxes}
The state-of-the-art Edge-Cloud platforms leverage the concept of sandboxes to provide isolation between the serverless functions. In the current container-based serverless platforms this is achieved with Linux \textit{namespaces} and \textit{cgroups}~\cite{Challenges-and-Opportunities}. Solutions such as SOCK~\cite{sock} and Faasm~\cite{faasm} leverage additional sandboxes to isolate functions, \textit{worker}, and \textit{faaslet}, respectively. Hence, functions in the same sandbox can safely share resources. While SOCK leverages IPC, Faasm uses distributed shared memory~\cite{faabric}; thus, co-located functions can benefit from proximity for low-latency inter-function communication. Although additional sandboxes decrease challenges such as cold start and inter-function latency, they add complexity and resource overhead with the additional sandboxes. 
SAND~\cite{sand} provides different isolation approaches based on the workflow. Containers isolate different workflows, while processes in the same container isolate functions within the same workflow. SAND enforces more robust isolation for different workflows and leverages container sharing in the same workflow to mitigate cold start issues. Additionally, shared libraries within the workflow must only be loaded once per workflow. Further, SAND's workflow container sharing decreases memory footprint as new functions are isolated by processes instead of isolated containers. Nevertheless, serverless functions are bound to SAND serverless framework to have these benefits decreasing portability and flexibility.
Sledge~\cite{sledge} leverages the Wasm sandbox mechanism to provide isolation, enabling untrusted modules to be safely executed on the host machine. As Sledge does not use standard container isolation, it is not OCI compliant, which means applications developed for Sledge are limited to a single Wasm runtime. CWASI follows OCI specifications to provide isolation and safety in addition to the Wasm sandbox. OCI specification enables compatibility, flexibility, and scalability within the state-of-the-art container manager and orchestration frameworks. Furthermore, CWASI selects the best communication approach based on function location and it optimizes co-located inter-function communication. 

\subsection{Third-party Services} 
Third-party services are the most common approach for inter-function data exchange, most specifically remote storage~\cite{why-when-how}. SAND~\cite{sand} introduces a hierarchical message bus to provide low latency with a global and a local bus. SAND optimizes co-located inter-function communication by introducing a local message bus responsible for exchanging data between functions on the same host. Nevertheless, SAND's message bus leverages Apache Kafka, which means SAND's approach relies on third-party services to exchange data even for co-located functions leading to unnecessary network traffic and infrastructure overhead since additional systems require additional infrastructure and operational effort. SONIC~\cite{sonic} offers three different approaches for inter-function communication: 
    \begin{enumerate*} [(i)]
        \item host storage where co-located functions leverage local storage to exchange data;
        \item direct passing where functions in different hosts exchange data by sending to the next function a data reference, e.g., IP. and file path;
        \item remote storage by using third-party services such as S3 and MinIO.
    \end{enumerate*}
SONIC provides low latency and flexibility by selecting the most suitable approach based on the functions and data location. 
CloudBurst ~\cite{cloudburstSF} relies on a local cache on each function host to allow low latency access to frequent data from a remote KVS Anna~\cite{anna}. Although Anna is low latency and highly scalable KVS, it proposes data exchange via remote third-party service. Regardless of storage or KVS, SONIC and CloudBurst still rely on third-party services to exchange data leading to additional network, resource, and data overhead which leads to duplicated data serialization.
CWASI introduces an optimized co-located inter-function communication that leverages IPC to exchange data directly through the host kernel memory buffer, decreasing latency and increasing throughput significantly. Additionally, CWASI enables Wasm static-linking to enable modules reuse and shared memory between Wasm modules decreasing latency.

\subsection{Co-located Inter-function Communication}
Co-located functions that rely on third-party services to exchange data do not profit from the function's placement to exchange data. 
SAND~\cite{sand} creates a local bus for co-located inter-function communication. Nevertheless, it relies on a third-party service to transfer the data, i.e., Apache Kafka.
ExCamera ~\cite{excamera} proposes a long-lived rendezvous server that receives and forwards requests between source and target workers. As inter-function relies on an external component, it adds unnecessary network requests similar to the third-party service approaches.
Nigthcore~\cite{nightcore} uses shared memory to exchange data and Linux pipes as inter-process communication to notify when data is available for reading or writing in the memory region. Nightcore provides microseconds latency but requires functions to include its runtime library, limiting portability.
POCKET~\cite{pocket} enables shared applications with cross-clients by creating an IPC channel with shared memory for low latency and high isolation. Floki~\cite{floki} leverages sync pipes for co-located inter-function communication and TCP sockets for remote communication. Floki provides low-latency communication by skipping multiple network communication layers and connecting directly to the POSIX layer. Nevertheless, Floki relies on an additional component forwarding agent deployed in the host namespace to create and connect to the TCP socket and to forward data between two functions. As Floki's forwarding agent is an extra component deployed via orchestration tools such as Kubernetes, it means that to be scalable; the forwarding agent needs to be separately scaled in addition to the function, which causes resource overhead in data-intensive scenarios.
CWASI introduces a container runtime shim to provide optimized co-located inter-function communication via a local buffer. Since CWASI is a shim, it lives along the function process, avoiding dependencies on additional components. Moreover, CWASI enables seamless Wasm static-linking between modules. Wasm-based serverless workflows with multiple modules run on a single Wasm VM, decreasing latency and memory footprint while increasing Wasm modules' reusability.

\section{CWASI Model and Architecture Overview}
\label{sec4}

\subsection{CWASI Model Overview}

\begin{figure}[!ht]
\centering
\hspace*{-2mm}
\includegraphics[width=.47\textwidth]{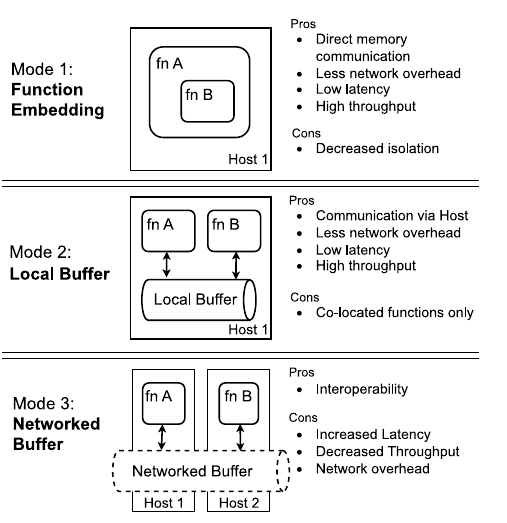}
\caption{CWASI Inter-function Communication Model}
\label{fig:comm_model}
\end{figure}

CWASI introduces a novel model for efficient inter-function communication between serverless functions. The model exploits function locality, i.e., functions co-located on the same host. 
\cref{fig:comm_model} shows the CWASI inter-function three-mode communication model featuring three modes: Function Embedding, Local Buffer, and Networked Buffer. 
Function Embedding communication model proposes one sandbox for trusted serverless functions. Local Buffer optimizes data exchange through the host resources, while Networked Buffer uses state-of-the-art mechanisms to communicate remote serverless functions. CWASI communication model integrates automatic inter-function communication selection and autonomous provision. CWASI enables inter-function communication by identifying, selecting, and provisioning the most suitable communication mode for each function. Consequently, it allows functions to have an optimized data exchange without external intervention. Next, we discuss the main model abstractions in more detail.

\paragraph{Function Embedding} This communication mode groups trusted functions into one sandbox, allowing them to share the same resources.
Thus, these functions can directly call each other via volatile memory such as shared memory, heap memory, and registers, leading to high-speed data access. Therefore, this communication mode is the most efficient. Function Embedding relies on CWASI to embed different functions in one sandbox. Thus, it decreases the function isolation. Nevertheless, functions in the same application namespace are considered trusted and do not need strong isolation~\cite{sand,webAssembly-in-non-web}. In CWASI shim, we leverage Wasm static-linking to embed tightly coupled and fully trusted Wasm functions into one single Wasm VM.

\paragraph{Local Buffer} CWASI leverages the host mechanisms to create a Local Buffer that enables co-located functions to exchange data. Hence, co-located functions can communicate without external services or network communication. This communication mode keeps different functions in its dedicated sandbox, enforcing a more robust isolation than Function Embedding. Therefore, the Local Buffer communication mode enables CWASI to execute functions that require higher trust, e.g., functions from different namespaces. During function startup, CWASI detects functions on the same host and creates a Local Buffer for each function.

\paragraph{Networked Buffer} Networked Buffer communication extends CWASI capabilities beyond co-located functions, allowing inter-function communication in distributed environments. Network communication is the standard communication between serverless functions. Thus, it provides a simplified deployment without any additional software required. Nevertheless, it increases latency, throughput, and network overhead.

\paragraph{Locality awareness and model selection} CWASI inter-function communication model improves data exchange with a mode selection that is transparent to the functions. To achieve this, the communication model must determine the function locality. CWASI model decides between Function Embedding and Local buffer for co-located functions based on the function trust level. For instance, functions in the same namespace are trusted and, therefore, eligible for Function Embedding. In contrast, functions in different namespaces are not trusted thus, only eligible for Local Buffer communication mode. Additionally, functions can give hints via deployment annotations to allow the CWASI model to choose one specific communication mode. Once the CWASI model has selected the best communication mode, it automatically provisions it.


\subsection{CWASI Shim Architecture Overview}

CWASI is a container runtime shim that mediates the communication of the container manager with the container runtime. For isolation, it relies on \textit{cgroups} and \textit{namespaces} on the host machine. Additionally, CWASI leverages WasmEdge runtime to create an isolated Wasm VM. Wasm deny-by-default mode provides security by only allowing explicitly requested host access via WASI~\cite{faasm, wasi}. CWASI proposes a novel design while maintaining interoperability with standard container managers and Wasm shims, which means CWASI runs alongside existing Wasm shims, e.g., RunWasi shims.

\begin{figure}[!ht]
\centering
\includegraphics[width=.49\textwidth]{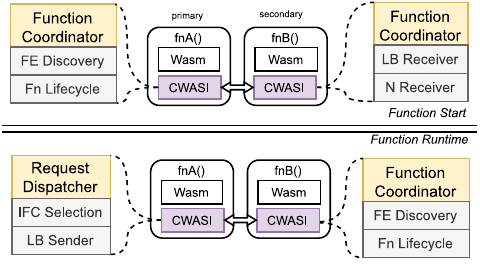}
\caption{CWASI Shim Architecture Overview}
\label{fig:cwasi_arch}
\end{figure}

\cref{fig:cwasi_arch} shows how CWASI interacts with wasm-based serverless functions during startup and function runtime to enable three-mode inter-function communication. We label functions as primary and secondary according to the workflow execution order. For example, given a workflow composed of fnA and fnB, where fnA invokes fnB. In this example, fnA is a primary, while fnB is a secondary function. During the function start, CWASI decides whether to the first, second, or third mode of the CWASI communication model.
At function runtime, CWASI fnA acts as a forward proxy identifying the best inter-function communication option and forwarding the requests. Whereas CWASI fnB acts as a reverse proxy receiving requests and initiating fnB function startup. 

\cref{fig:cwasi_arch} shows CWASI core components: \textit{Function Coordinator} and  \textit{Request Dispatcher}.
These core components implement the following features:
\begin{enumerate*} [label=(\roman*)] 
\item \textit{Function (Fn) Lifecycle};
\item \textit{Function Embedding (FE) Discovery}; 
\item \textit{Local Buffer (LB) Receiver}; 
\item \textit{Network (N) Receiver};
\item \textit{Inter-Function Communication (IFC) Selection};  and
\item \textit{Local Buffer (LB) Sender}.

\end{enumerate*}

\paragraph{\textit{Function Coordinator:}} It is a core component responsible to start and maintain the function. This component is an enabler for the three-mode communication. During function start, the function coordinator starts \textit{Function Embedding Discovery} for the first communication mode, \textit{Local Buffer Receiver} for the second, and \textit{Network Receiver} for the third, as displayed in \cref{fig:cwasi_arch}. In addition to the communication-related tasks, the function coordinator manages the \textit{Wasm Lifecycle}. For the task decision-making process, the function coordinator relies on the OCI bundle annotations, as described in \cref{alg:flc_start}.
 

In \cref{alg:flc_start}, let $O$ be the OCI spec which consists of arguments $A_i$, annotations $N_j$, where $i$ and $j$ are indexes that represent the elements in the set $A_i$ and $N_j$, a bundle path $B$, and remaining fields represented by $F$. Let $S$ be a string identifier. If the identifier $S$ exists in the set of annotations $N$, then \cref{alg:flc_start} starts a \textit{Network Receiver} with the function name, where the function name is the first element $A_0$ of the argument set $A$. Further, it creates a \textit{Local Buffer Receiver} for function name $A_0$ and bundle path $B$. If $S$ does not exist in $N$, then the manager writes Wasm binary file to WAT, loads the discovered Wasm function modules into Wasm VM function and finally triggers \textit{Wasm Lifecycle} which will start the Wasm VM function with the set of arguments $A$.
\begin{algorithm}[!htb]
\caption{Function Coordinator}
\label{alg:flc_start}
\begin{algorithmic}[1]
\Require O = ($\text{A}_i, N_j, B, F$) where $i,j \in \mathbb{N}_0$
 : oci spec
\State S: string primary/secondary identifier 
\If {$ \exists S \in N$}
     \State $network\_receiver(A_0)$
     \State $kernelbuffer\_receiver(A_0,B)$
\Else 
    \State $wat \gets wasmprinter(B)$
    \State $wasm\_modules \gets function\_embedding\_discovery(wat)$
    \State $embedd\_modules\_in\_function(wasm\_modules)$
    \State $wasm\_lifecycle(A)$ 
\EndIf
\end{algorithmic}
\end{algorithm} 

    \begin{enumerate} [wide=10pt,label=(\roman*)]
        \item \textit{Function Lifecycle:} This feature is responsible for function lifecycle methods such as \path{new}, \path{start}, \path{wait}, \path{delete}, \path{kill}. This feature creates and executes Wasm VM and Linux processes. Once the function has finished, it terminates every process related to the specific function. Additionally, it connects directly with the container manager to give updates on the container status.
        \item \textit{Function Embedding Discovery:} This feature discovers whether tightly coupled additional modules are present on the machine. If such a module is found, it embeds these additional modules in the same Wasm VM function. Hence, during runtime these Wasm modules share the same memory stack, allowing them to communicate directly in the first mode of the CWASI communication model. 
        \item \textit{Local Buffer Receiver:} It is a feature that creates a Local Buffer Receiver based on the functions OCI annotations. 
        CWASI shim creates this after identifying whether the serverless function is primary or secondary, e.g., if \textit{fnA} connects to \textit{fnB}, \textit{fnA} does not need a Local Buffer Receiver but \textit{fnB} does. Therefore, we label \textit{fnB} as secondary, and based on the annotations on the OCI spec, the shim can decide whether to create a Local Buffer Receiver and a Network Receiver for this function or not.
        \item \textit{Network Receiver:} This feature enables the secondary function to receive messages via network communication. In CWASI, we enable network communication with message queues. During the function start, the function coordinator creates a dedicated queue for each function which will during runtime accept incoming messages via network.
    \end{enumerate}
\paragraph{\textit{Request Dispatcher:}} This core component identifies during runtime which communication model is more efficient for the task at hand, according to the three modes from the communication model presented in \cref{sec4}. 

    \begin{enumerate} [wide=10pt,label=(\roman*),start=5]
        \item \textit{Inter-Function Communication (IFC) selection:} Triggered by \textit{Request Dispatcher} component, this feature identifies during runtime whether running serverless functions are co-located or remotely. CWASI Shim leverages the container manager to learn which functions run on the same host. Further, either it selects a target function on the same host, or it sends the request via the Networked Buffer if the function is placed on a different host.
    
        \cref{alg:ifc_selection} shows how CWASI determines whether the functions run locally or remotely during runtime. In \cref{alg:ifc_selection}, $FT$ is the input function target type $FT$ e.g. \textit{fnB}, and container manager running path $RP$, while local buffer receiver path $SP$ is the output. \cref{alg:ifc_selection} iterates every function path $F_P$ of OCI spec file $F$ in running path $RP$. In each interaction, the function reads the OCI config for file path $F_P$. Let $C$ be an object that is composed of a set of arguments $A$ and remaining fields $RF$ obtained from $read\_oci\_config$. If there exists target function $FT$ in argument set $A$ then, set local buffer receiver path $SP$ is equal to the function path $F_P$ plus $S$, where $S$ is a string local buffer receiver suffix.
        
        \begin{algorithm}[!htb]
        \caption{IFC Selection}
        \label{alg:ifc_selection}
        \begin{algorithmic}[1]
        \Require $FT$, $RP$: function type and running path
        \Ensure  $SP$: local buffer receiver path
        \ForAll { $ F_P \in F, \forall F \in RP $ }
            \State $C = (A,RF), \text{where } C \gets read\_oci\_config(F_P)$
            \If {$\exists FT \in A$}
                \State \Return $SP \gets F_P + S$
            \EndIf
        \EndFor
        \end{algorithmic}
        \end{algorithm}
        \item \textit{Local Buffer Sender:} After identifying a serverless function on the same host, the \textit{Request Dispatcher} component triggers this feature during runtime. The Local Buffer Sender enables the shim to communicate with a Local Buffer Receiver from another serverless function shim on the same host synchronously. It is responsible for halting the shim until a response from the Local Buffer Receiver arrives.
    \end{enumerate}

CWASI inter-function communication model displayed \cref{fig:comm_model}, and its architecture shim, in \cref{fig:cwasi_arch}, address RC-1. First, we present a model for inter-function communication. Further, we describe the CWASI architecture that enables this communication model.



\section{CWASI Three-mode inter-function communication} \label{sec5}
CWASI introduces three key mechanisms for efficient and low-latency data exchange for serverless workflows in the Edge-Cloud continuum. These mechanisms enable each mode from the communication model presented in ~\cref{sec4}. CWASI facilitates function embedding by statically linking Wasm functions from the same namespace. In the second mode, CWASI leverages Unix Sockets to enable local buffer communication for functions in different namespaces. In the third mode, CWASI enables networked buffer communication via state-of-the-art mechanisms.

\subsection{Function Embedding Communication Mode}

For function embedding, CWASI enables Wasm functions static linking by analyzing the WAT file. Once the shim knows every necessary import for a specific function, it leverages the container manager snapshot to identify if any required import exists in the host~\cite{static_link}. This mechanism is executed before the \path{start} function part of the \textit{Wasm Lifecycle} feature of the \textit{Function Coordinator} component. Whenever the Wasm functions start, the shim is aware of every statically linked function enabling a seamless Wasm function integration with low latency communication.

To achieve that, before starting \textit{fnA.wasm}, CWASI reads the \textit{fnA.wasm} in WAT and searches for imports. If such import exists on the host, then \textit{fn-utils.wasm} can be loaded into the same Wasm VM. CWASI statically links \textit{fn-utils.wasm} with \textit{fnA.wasm} Wasm VM during VM instantiation. Static links enable \textit{fnA.wasm} to reuse  \textit{fn-utils.wasm} function without creating a second Wasm VM. Additionally, during execution time, \textit{fnA.wasm} and \textit{fn-utils.wasm} share the same Wasm VM memory address facilitating the data exchange between the Wasm binaries~\cite{static_link}. 
Once statically linked, \textit{fnA.wasm} leverages FFI mechanisms to call method \path{get_image_metatada(image_file)} in \textit{fn-utils.wasm}. In this specific situation, the communication remains within the same Wasm VM. \textit{fnA.wasm} can access the memory from \textit{fn-utils.wasm} to exchange data enabling the communication between the two functions as one function.
\cref{alg:wim_detection} shows how the \textit{Function Embedding Discovery} detects additional Wasm functions import in the WAT file to enable Wasm static linking. Further, the \textit{Function Embedding Discovery} looks for the bundles that match the WAT imports in the container manager snapshot, i.e., \textit{containerd} snapshot, and returns them to the \textit{Function Coordinator}. If the imported function exists in the snapshot path, the \textit{Function Coordinator} loads these functions into the Wasm VM before starting it, as described in \cref{alg:flc_start}.

In \cref{alg:wim_detection}, let the WAT file text be input $W$ and $B$, the result set containing all bundles' matches encountered by the \cref{alg:wim_detection}. For every file path $F_P$ where $F_P$ belongs to a single file object $O$. $O$ is part of a set of files $F$ in the container manager snapshot. $P$ is a set of import pattern matches in the WAT file $W$. If there exists an element $P$ such that file path $F_P$ equals $P$, then add $F_P$ to the result set $B$.
The result set $B$ is returned to the \textit{Function Coordinator}, which statically links the functions in the Wasm VM. \cref{alg:flc_start} receives input $VM$ API provided by the Wasm runtime and a set $M$ with the functions returned from \cref{alg:wim_detection}. Line 9 starts the Wasm VM with every function $m$ in set $M$ to the Wasm $VM$. Once every function is registered, they can seamlessly communicate as one application.

\begin{algorithm}[!htb]
\caption{Function Embedding Discovery}
\label{alg:wim_detection}
\begin{algorithmic}[1]
\Require $W$: WAT file text 
\Ensure  $B$: bundles path set
\State $B \gets \emptyset$
\State $F$ $\leftarrow$ files in container manager snapshot
\ForAll { $F_P (F_P \in O, \forall O \in F)$ }
    \If{ $\exists P (F_P = P \land P \in W)$ }
        \State $B \cup F_P$
    \EndIf
\EndFor
\end{algorithmic}
\end{algorithm}

By enabling Function Embedding via Wasm static linking, CWASI addresses the research challenge RC-2, described in ~\cref{sec2}. It enables multiple Wasm functions to share the same address space, facilitating more efficient communication since the data can be transferred directly from the memory. Although loading multiple Wasm functions into a single Wasm VM directly affects the Wasm VM isolation, CWASI considers these functions to be fully trusted code that belongs to the same serverless function. Thus, these Wasm functions do not require to be executed in separate Wasm VMs.

\subsection{Local Buffer Communication Mode}\label{mec:co-located}

CWASI reduces the network traffic by forwarding the local traffic via a local buffer instead of using the network for local communications. By enabling local buffer communication, CWASI addresses the research challenge RC-3, described in ~\cref{sec2}.

\begin{figure}[!ht]
\centering
\includegraphics[width=0.35\textwidth]{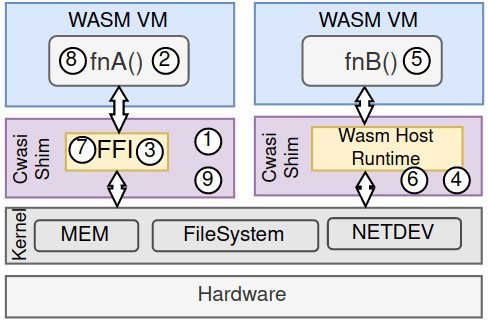}
\caption{Local Buffer Inter-function Communication}
\label{fig:local_func}
\end{figure}

\cref{fig:local_func} shows that \textit{fnA} and \textit{fnB} run on their own isolated Wasm VM, and each of them has its own CWASI shim instance. Separated Wasm VMs are necessary to keep isolation between the functions. To enable functions to exchange data via a local buffer, we use a Foreign Function Interface (FFI) mechanism, where the Wasm binary executes a function outside of its Wasm VM but on its own CWASI shim process. 
First \textcircled{1} \textit{fnA} shim starts, it loads the Wasm function binary file. The binary file path is present on the OCI specification created by the container manager. Then \textit{fnA} shim executes \textit{fnA} function;
\textcircled{2} \textit{FnA} function triggers Request Dispatcher on \textit{FnA} shim which is executed with a pointer reference and data length from \textit{fnA} input data;
\textcircled{3} The Request Dispatcher on \textit{fnA} shim
looks for running \textit{fnB} in the container manager, and in positive cases connects to \textit{fnB} socket and sends \textit{fnA} function input;
\textcircled{4} \textit{FnB} shim accepts \textit{fnA} connection request, creates and executes Wasm \textit{fnB} function;
\textcircled{5} \textit{FnB} function is executed and its result returned to \textit{fnB} shim;
\textcircled{6} \textit{FnB} shim receives the function result and responds to \textit{fnA} shim. Afterward, it starts its shutdown, including closing the server socket;
\textcircled{7} \textit{FnA} shim receives the input from \textit{fnB} shim and writes the result into the \textit{fnA} function memory; 
\textcircled{8} \textit{FnA} read the memory with pointer and size returned from Request Dispatcher and resumes its execution; 
\textcircled{9} Once \textit{fnA} has finished, \textit{fnA} shim returns the output to the container manager and starts its shutdown.

CWASI provides the Request Dispatcher to enable data exchange between functions during runtime, as shown in \cref{fig:local_func}. Request Dispatcher is a host function on the shim that enables host runtimes to communicate with Wasm functions via imports and exports. The Wasm function triggers the Request Dispatcher via an FFI that connects the Wasm function to the host function on the shim. 
As shown in \cref{alg:ffi_external_func}, the \textit{Request Dispatcher} receives a memory API as input. The memory API gives access to Wasm VM, enabling \textit{Request Dispatcher} to read and write directly into Wasm VM memory.
Once the \textit{Request Dispatcher} reads the input data from the source serverless function, it serializes the input and extracts the source and target function to create an inter-function communication pair. With the container manager running path from the OCI configuration file, the \textit{Request Dispatcher} triggers the \textit{ifc\_selection}, shown in \cref{alg:ifc_selection}, to look for the target function. 

In \cref{alg:ffi_external_func}, let memory API $M$ and input values set $V_i$ to be the input. Let the output be a result string reference $R$, where $R$ is composed by pointer $R_P$ and string length $R_L$. Let function type $FT$ and function source payload $PL$ be part of the string read via memory API $M$. 
\cref{alg:ffi_external_func} reads $FT$ and $PL$ via $M$ with the first two elements from input $V$, pointer $V_0$ and length $V_1$. $V_0$ and $V_1$ compose a reference to a string in the Wasm VM.
$RP$ is the container manager running path statically accessible to the shim. $SP$ is the result from $ifc\_selection$ which receives target function type $FT$ and running path $RP$ as input. In cases where socket path $SP$ is not empty, \cref{alg:ffi_external_func} connects to the target socket path $SP$ and sends function source payload $PL$. When socket path $SP$ is empty, the function publishes payload $PL$ into the target function type queue $FT$. Let the output of $connect\_socket$ and $publish\_queue$ be a string with a pointer reference and string length. Then, \cref{alg:ffi_external_func} writes into the Wasm VM memory $M$ the result string $R$ using result pointer $R_P$ and result string length $R_L$. Finally, the function returns $R$ containing results pointer reference $R_P$ and length $R_L$.

\begin{algorithm}[!htb]
\caption{Request Dispatcher}
\label{alg:ffi_external_func}
\begin{algorithmic}[1]
\Require $M,V_i$ where $i \in \mathbb{N}_0 $: VM memory API and input set 
\Ensure  $R = (R_P, R_L)$ : Result pointer and length
\State FT, PL $\in$ M.read($V_0$,$V_1$)
\State $RP \gets $ current container manager running path
\State $SP$ $\gets ifc\_selection(FT, RP)$ 

\If{$SP \ne \emptyset$}
    \State $R \gets connect\_socket(SP,PL)$
\Else
    \State $R \gets publish\_queue(FT,PL)$
\EndIf
\State $M.write(R_P,R_L)$
\State \Return $R$
\end{algorithmic}
\end{algorithm}

\subsection{Networked Buffer Communication Mode}
On the third mode, CWASI enables inter-function communication by leveraging existing state-of-the-art publish/subscribe mechanisms instead of relying on typical storage solutions. Direct inter-function communication leads to significant latency and throughput improvement when compared to solutions that rely on remote services such as remote storage, KVS, and databases~\cite{sand}.
\section{CWASI Implementation} \label{sec6}

CWASI is published as an open-source shim part of the Polaris SLO Cloud. Polaris itself is part of the Linux Foundation Centaurus project. It is implemented in Rust and currently supports WasmEdge runtime.  source code is available on GitHub. It is implemented in Rust and currently supports WasmEdge runtime. CWASI source code is available on GitHub\footnote{\url{ https://github.com/polaris-slo-cloud/containerd-shim-cwasi}}.
We are influenced by RunWasi, which introduces runtime shims for the state-of-the-art Wasm runtimes WasmEdge and Wasmtime. For this implementation, we connect the container manager \textit{containerd} using the protobuf provided by RunWasi crates.
CWASI adopts lifecycle-related methods from RunWasi, meaning they connect directly to \textit{containerd} via protobuf.  In the Wasm context, the shim is a \textit{host runtime} that interacts with Wasm binaries via imports and exports. Currently, CWASI supports WasmEdge and leverages WasmEdge SDK~\cite{Wasmedge} to interact with Wasm binaries' memory space to exchange data between different modules. We have chosen WasmEdge runtime for the variety of its available features and community support.

\paragraph{Inter-function communication} For local buffer communication mode, CWASI leverages IPC – Unix Domain Socket (UDS) for inter-function data exchange on the same host to enable bi-directional data exchange between the functions. Unlike other IPC communication mechanisms such as Unix pipe, UDS allows the communication between two non-related processes, i.e., processes without direct communication or co-dependencies to exchange data\cite{pipe}. For co-located inter-function communication between two shims, we leverage UDS, creating a temporary file with OCI bundle name and \path{.sock} suffix in the container manager snapshot, e.g., \path{/var/run/containerd/io.containerd.runtime.v2.task/mycontainer.sock}. The shim shuts down every socket server and removes the temporary socket file when the process finishes. Alternatively, in case of errors, the container manager sends a kill signal to shim, which cleans up the container process resources, shutdown the socket, and removes the socket file. We leverage the rust \path{UnixListener} to create a socket server type \path{AF_UNIX} and \path{UnixStream} to create a socket client from rust module \path{std::os::unix::net}\footnote{\url{https://doc.rust-lang.org/std/os/unix/net/index.html}}. For remote inter-function communication, we leverage Redis Pub/Sub\footnote{\url{https://redis.com/solutions/use-cases/messaging}}. 

\paragraph{Complex data transfer} Currently WebAssembly only supports numbers as data types, i.e., integers and float of 32-bit and 64-bit each. Therefore transferring complex data such as strings is a challenging task~\cite{webassembly-specs}. As CWASI is a host runtime with access to the Wasm VM memory space, we leverage WasmEdge APIs to transfer complex data to the shim such as string by sending a pointer reference along with the data length. Further, the shim reads the byte array from memory and converts it to a string. The shim does the reverse action to enable Wasm modules to receive strings during runtime. It writes the byte array into the Wasm VM memory and returns the pointer and length to the Wasm module. Finally, the Wasm modules can retrieve the byte array to a string. Libraries such as wasm-bindgen and wasmedge-bindgen attempt to facilitate a high-level complex data transfer interaction between Wasm modules. Nevertheless, wasm-bindgen cannot be applied for data transfer between the Wasm module and host function until the current moment of writing this paper.

\section{Experiments} \label{sec7}

\subsection{Overview}

\begin{figure}[!htb]
\centering
    \begin{subfigure}[t]{0.12\textwidth}
    \centering
    \vspace*{-12.5mm}
    \includegraphics[width=\textwidth]{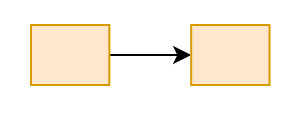}
    \vspace*{0.4mm}
    \caption{Sequential} \label{fig:seq}
\end{subfigure}
\begin{subfigure}[t]{0.12\textwidth}
    \centering
    \includegraphics[width=\textwidth]{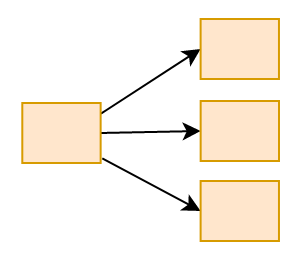}
    \caption{Fan-out} \label{fig:fanout}
\end{subfigure}
 \begin{subfigure}[t]{0.12\textwidth}
    \centering
    \includegraphics[width=\textwidth]{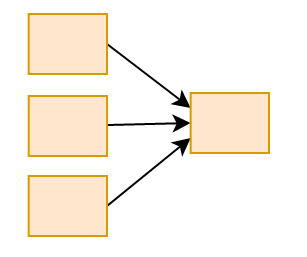}
    \caption{Fan-in} \label{fig:fanin}
\end{subfigure}

\caption{Experimental Workflows}
\label{fig:workflows}
\end{figure}

We design experiments to evaluate our scenario application using Serverless workflow's most important invocation patterns, including \textit{Sequential}, \textit{Fan-out}, and \textit{Fan-in} (\cref{fig:workflows}), as outlined in~\cite{jonas2019cloud}. Furthermore, the experiments aim to measure the performance of the main contributions of this paper, listed in ~\cref{sec1}.

\cref{fig:seq} shows an example of a \textit{Sequential} workflow when only one instance of \textit{Extract Frames}, \textit{Process Frames}, and \textit{Prepare Dataset} is created, so each function is called sequentially. A \textit{Fan-out} workflow, shown in \cref{fig:fanout}, happens when one function triggers multiple parallel functions. In our scenario, that occurs when the resulting frame from \textit{Extrac Frames} stage triggers multiple \textit{Process Frames} e.g., one function for labeling and another function for anonymization. Finally, we have a \textit{Fan-in}, shown in \cref{fig:fanin}, when multiple parallel functions trigger one single function. In our scenario, that happens when multiple instances of \textit{Process Frames} functions trigger one function \textit{Prepare Dataset}. We describe the results of evaluating these workflow composition patterns in \cref{exp:seq}, \cref{exp:fanout}, and \cref{exp:fanin}, respectively.

We compare CWASI to the baselines Runwasi WasmEdge shim~\cite{runwasi} and a container solution OpenFaas~\cite{openfaas}. CWASI Shim focuses on inter-function communication. Therefore, we measure the specific time frame between two serverless functions, when the shim sends the message until the subsequent shim receives the message. We use an HTTP client and server for WasmEdge for inter-function communication. For OpenFaas, we use its standard API Gateway, which is also an HTTP server client solution.

One of the main characteristics of Serverless Computing is the function composition between small stateless short-lived functions. As stateless functions, inter-function data exchange is crucial for serverless computing~\cite{serverlessbench}. We must ensure that functions can communicate on a large scale with low latency without adding extra resource overhead on the multi-tenant environment from the Edge-Cloud continuum. Hence, we perform experiments to show performance, scalability, and potential drawbacks in resource usage. For this purpose, we collect the following metrics:

\paragraph{Latency} This metric shows the exact execution time when \textit{fnA} shim sends the message until \textit{fnB} shim receives the message. We use seconds as the unit metric for the latency experiments.
\paragraph{Throughput} With this metric, we want to examine if the optimization performed does not affect the scalability of serverless functions under high load. We use requests per second to indicate the throughput. In cases where the executions are lower than one second, we extrapolate, e.g., if we send ten requests that take less than one second, we extrapolate the throughput by considering the rate of execution over a one-second timeframe.
\paragraph{Resource Usage} As our co-located inter-function leverages the host resources, we measure the resource usage of the host machine to show that CWASI shim performs similarly or at maximum small increases compared to the baseline metrics. We display the CPU usage in \% and the RAM usage in Kb.

\subsection{Experiment Setup}

To evaluate CWASI shim, we execute the designed workflows with CWASI and, in the baselines WasmEdge and OpenFaas. Our Serverless functions for co-located experiments are performed in a single machine Intel NUC with i5 2.9 GHz 8GB RAM with Ubuntu 22.04 LTS. This machine executes all serverless functions. We perform OpenFaas experiments with Openfaas faasd~\cite{openfaas}. As container manager, we use containerd and its client cli tool \footnote{\url{https://github.com/projectatomic/containerd/blob/master/docs/cli.md}} to execute the serverless functions.  Additionally, we use shell scripts to start multiple functions for every baseline. To avoid biased results, we repeated the experiments ten times and collected the average result.

\subsection{Sequential Workflow}\label{exp:seq}

\begin{figure*}[!ht]
\centering
    \begin{subfigure}[t]{0.33\textwidth}
    \centering
    \includegraphics[width=\textwidth]{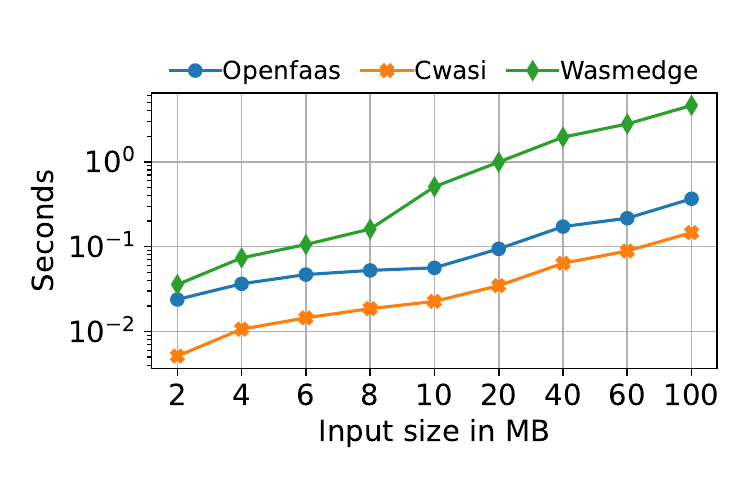}
    \caption{Latency} \label{fig:seq_result_latency}
\end{subfigure}
\begin{subfigure}[t]{0.33\textwidth}
    \centering
    \includegraphics[width=\textwidth]{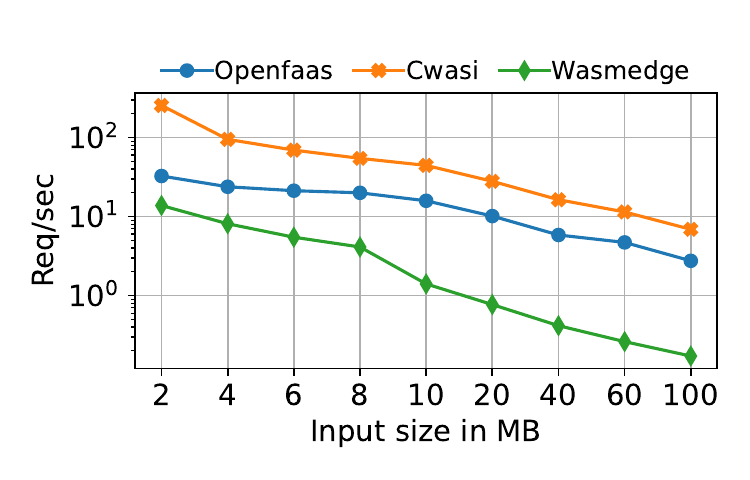}
    \caption{Throughput} \label{fig:seq_result_throughtput}
\end{subfigure}
 \begin{subfigure}[t]{0.33\textwidth}
    \centering
    \includegraphics[width=\textwidth]{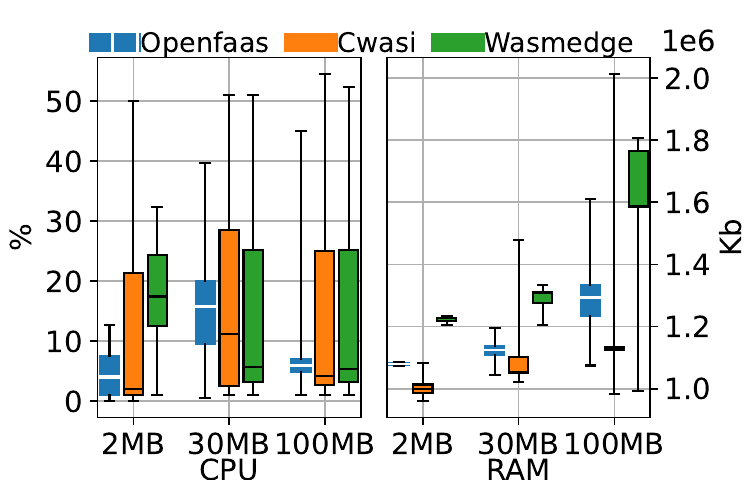}
    \caption{Resource} \label{fig:seq_result_resource}
\end{subfigure}

\caption{Sequential Results}
\label{fig:seq_result}
\end{figure*}

In this experiment, we perform synchronous requests between two serverless functions fnA and fnB as shown in \cref{fig:seq}. We increase the input size during the experiment to evaluate the solution with different loads. \cref{fig:seq_result} shows overall latency, throughput, and resource results while \cref{tab:seq_result} shows precise measurements extracted from the overall results. 

\cref{fig:seq_result_latency} displays the input data size in MB on the $x$-axis and the latency on the $y$-axis. Over input size increase, CWASI shows from 5 to 143 ms, WasmEdge shows from 48 ms to 4.4 s, while OpenFaas shows from 24 ms to 368 ms. The latency results show CWASI decreases up to 79\% and up to 89\% of the latency when compared to OpenFaas and WasmEdge, respectively. The expected linear increase shows application stability in all three cases with consistent performance.

\cref{fig:seq_result_throughtput} shows the throughput of the systems. $x$ axis shows the input data size and the $y$ axis shows how many requests the systems can handle per second. The results show a linear throughput decrease over the input size increase, which means the complexity is at maximum linear. According to \cref{tab:seq_result}, CWASI shows a throughput from 266 to 6.9, WasmEdge from 15 to 0.17, and OpenFaas from 41 to 2.7 requests per second. The results show CWASI increases the throughput up to 5x when compared to OpenFaas and up to 14x compared to WasmEdge.

\cref{fig:seq_result_resource} shows RAM and CPU usage for CWASI, OpenFaas, and WasmEdge. We selected three input sizes representing the small, mid, and large input sizes with latency and throughput experiments. On the left side of \cref{fig:seq_result_resource}, we notice a higher CPU usage from CWASI and WasmEdge. As OpenFaas only includes the scale to zero feature in the pro bundle and we perform the experiments with the community version, our experiments with OpenFaas have two functions running constantly. CWASI and WasmEdge had different functions and processes for each request in this experiment. The slight difference in the experiment design does not affect the latency and throughput but leads to lower CPU usage as the functions are reused in OpenFaas. The right side of \cref{fig:seq_result_resource} shows RAM usage. We observe a decrease of up to 30\%  usage in CWASI compared to WasmEdge and up to 15\%  compared to OpenFaas.

\begin{table}[!htb]
\caption{Sequential Result}
\begin{center}
\begin{tabular}{|c|c|c|c|c|}
\hline
\multicolumn{5}{|c|}{\cellcolor[HTML]{EFEFEF}\textbf{Sequential Workflow}}\\\hline
& \multicolumn{2}{c|}{\textbf{Latency sec}} & \multicolumn{2}{c|}{\textbf{Throughput Req/sec}} \\ \hline
    & \multicolumn{1}{c|}{2MB}   & 100M  & \multicolumn{1}{c|}{2MB}    & 100MB  \\ \hline
CWASI & \multicolumn{1}{c|}{0.0051}  & \multicolumn{1}{c|}{0.1436} & 266.4535 &  6.9569  \\ \hline
WasmEdge & \multicolumn{1}{c|}{0.048}  & \multicolumn{1}{c|}{4.4324} & 15.8323 & 0.1732   \\ \hline
OpenFaas & \multicolumn{1}{c|}{0.0248}   & \multicolumn{1}{c|}{0.3681} & 41.7240 & 2.7556    \\ \hline          
\end{tabular}
\label{tab:seq_result}
\end{center}
\end{table}

\subsection{Fan-out Workflow}\label{exp:fanout}

\begin{figure*}[!htb]
\centering
    \begin{subfigure}[t]{0.33\textwidth}
    \centering
    \includegraphics[width=\textwidth]{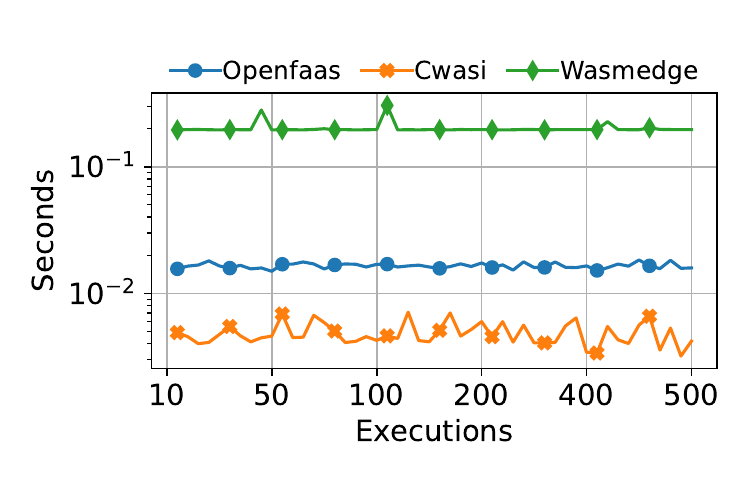}
    \caption{Latency} \label{fig:fanout_latency}
\end{subfigure}
\begin{subfigure}[t]{0.33\textwidth}
    \centering
    \includegraphics[width=\textwidth]{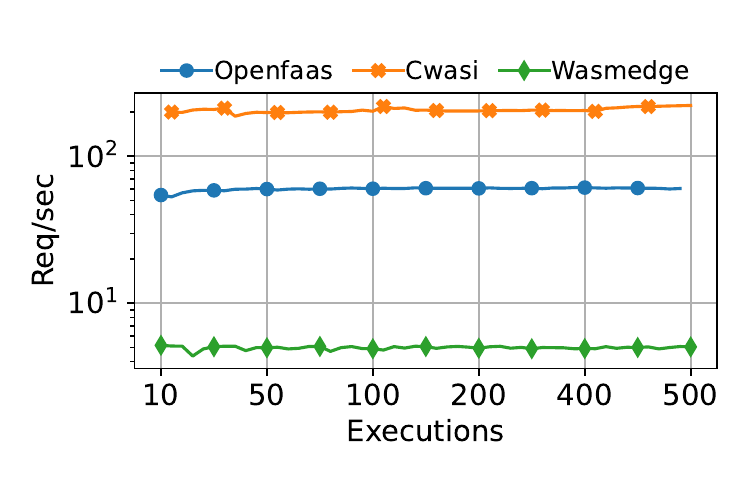}
    \caption{Throughput} \label{fig:fanout_throughput}
\end{subfigure}
 \begin{subfigure}[t]{0.33\textwidth}
    \centering
    \includegraphics[width=\textwidth]{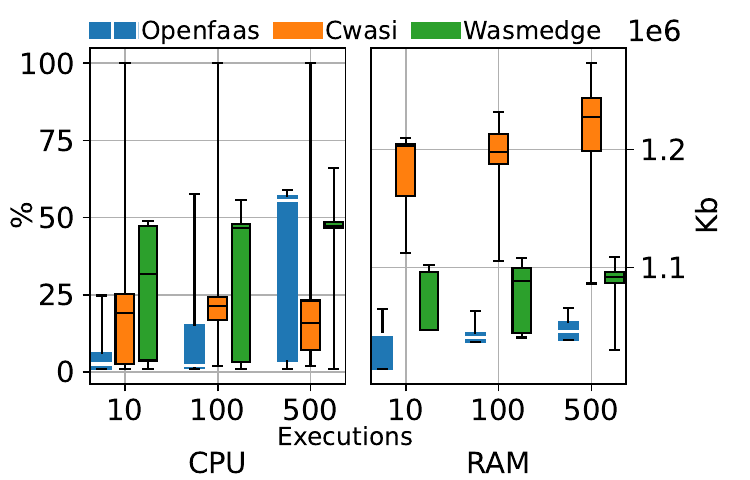}
    \caption{Resource} \label{fig:fanout_resource}
\end{subfigure}

\caption{Fan-out Results}
\label{fig:fanout_result}
\end{figure*}

In this experiment, we perform fan-out requests. As parallel requests are a regular part of real-world use cases, as described in \cref{sec2}, the goal of this experiment is to analyze the performance of CWASI shim compared to the baselines WasmEdge and OpenFaas when executing parallel requests. This experiment comprises one function that branches out and executes multiple parallel requests. Since WASI does not support multi-threading~\cite{wasm-multithreading}, we leverage wasmedge~\cite{Wasmedge} libraries such as hyper-wasi and tokio-wasi to execute Wasm asynchronous requests. 
\cref{fig:fanout} exemplifies the design of the fan-out experiments. We use a fixed input size of 2MB and increase the level of multiple parallel requests in the fan-out starting from 10 to 500 fan-out requests.

\cref{fig:fanout_result} shows latency, throughput, and resource usage results. In the latency results \cref{fig:fanout_latency}, we have the multiple parallel executions in axis $y$ and the duration in seconds in axis $x$. We observe stable horizontal lines with slight degrees of change for every three frameworks. \cref{tab2} shows a sample of metrics collected during this experiment. CWASI shows around 6 ms, WasmEdge shows 195 ms, and OpenFaas 16 ms. Although none of the frameworks display significant changes over the axis $x$, CWASI shows up to 62\% lower latency execution when compared to OpenFaas and up 95\% compared to WasmEdge.

Similar to \cref{fig:fanout_latency} latency results, we see horizontal lines representing the measured frameworks in the throughput results shown in \cref{fig:fanout_throughput}. On axis $x$, we have the fan-out degree, while on axis $y$ requests per second. The expected stable throughput for every framework indicates no performance issues under high load. According to \cref{tab2}, CWASI shows around 3 ms, WasmEdge 115 ms, and OpenFaas 7 ms. Hence, CWASI shows up to 1.3x higher throughput compared to OpenFaas and up to 37x compared to WasmEdge.

\cref{fig:fanout_resource} shows CPU and RAM usage for the fan-out experiment. On the left of \cref{fig:fanout_resource}, we have execution metrics samples on the axis $x$ and CPU percentage usage on the axis $y$. Overall, we observe a slight CPU usage increase between 10 and 500 executions where CWASI displays peaks of 100\% usage. On the right side of \cref{fig:fanout_resource}, we observe an increase in RAM usage from CWASI compared to OpenFaas and WasmEdge. The low resource usage is because WasmEdge and OpenFaas rely on the HTTP client and server where we could reuse the same HTTP server. At the same time, CWASI explicitly creates a new process for every new function, leading to higher CPU usage for CWASI. The processing creating increase reflects on the CPU usage peaks and RAM usage. The resource usage increase reflects latency and throughput results shown in \cref{fig:fanout_result} and in \cref{fig:fanout_throughput}, respectively.

\subsection{Fan-in Workflow}\label{exp:fanin}

In these experiments, we aim to measure CWASI scalability with multiple requests in fan-in workflows as shown in \cref{exp:fanin}. As CWASI assumes every request is a new function and consequently a new process, we have limitations to creating a fan-in workflow by simply calling the same function over and over. In this scenario, if a particular function is called ten times, CWASI creates ten different functions, which means fan-in is the same and ten different sequential workflows. To overcome this experiment limitation, we use the fan-out experiments and calculate the responses, revealing a fan-in workflow. Using the same set of experiments means the resource results shown in \cref{fig:fanout_resource} corresponds to fan-out and fan-in experiments. Thus, in this subsection, we address latency and throughput.

\begin{figure}[!htb]
\centering
    \begin{subfigure}[t]{0.33\textwidth}
    \centering
    \includegraphics[width=\textwidth]{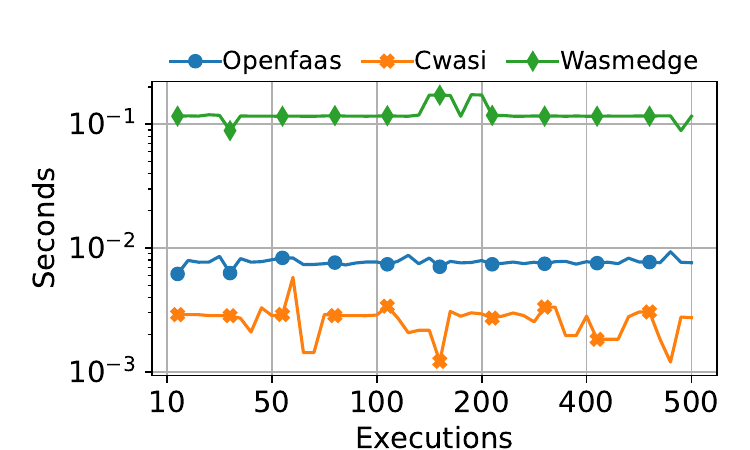}
    \caption{Latency} \label{fig:fanin_latency}
\end{subfigure}
\begin{subfigure}[t]{0.33\textwidth}
    \centering
    \includegraphics[width=\textwidth]{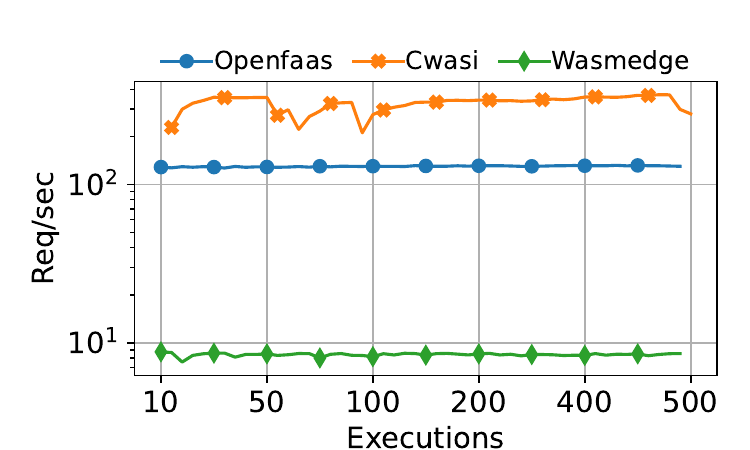}
    \caption{Throughput} \label{fig:fanin_throughput}
\end{subfigure}
\caption{Fan-in Results}
\label{fig:fanin_result}
\end{figure}

\begin{table}[!htb]
\caption{Fan-out and Fan-in Sample Results}
\begin{center}
\begin{tabular}{|c|c|c|c|c|}
\hline
\multicolumn{5}{|c|}{\cellcolor[HTML]{EFEFEF}\textbf{Fan-out Workflow}}\\\hline
& \multicolumn{2}{c|}{\textbf{Latency sec}} & \multicolumn{2}{c|}{\textbf{Throughput Req/sec}} \\ \hline
    & \multicolumn{1}{c|}{10 Exec}   & 100 Exec  & \multicolumn{1}{c|}{10 Exec}     & 100 Exec  \\ \hline
 
CWASI & \multicolumn{1}{c|}{0.0045} & \multicolumn{1}{c|}{0.0066}             & 203.7327 & 211.2039   \\ \hline
WasmEdge & \multicolumn{1}{c|}{0.1931} & \multicolumn{1}{c|}{0.1950}             & 4.3781 & 4.9118   \\ \hline
OpenFaas & \multicolumn{1}{c|}{0.0441} & \multicolumn{1}{c|}{0.0169}              & 59.8344 &  61.2589  \\ \hline
\multicolumn{5}{|c|}{\cellcolor[HTML]{EFEFEF}\textbf{Fan-in Workflow}}\\\hline
CWASI & \multicolumn{1}{c|}{0.0029} & \multicolumn{1}{c|}{0.0032}            & 298.9536 &  314.2019    \\ \hline
WasmEdge & \multicolumn{1}{c|}{0.1161} & \multicolumn{1}{c|}{0.1170}            & 8.6752 &   8.5336   \\ \hline
OpenFaas & \multicolumn{1}{c|}{0.0075} & \multicolumn{1}{c|}{0.0079}             & 127.3925 &  130.8297    \\ \hline
\end{tabular}
\label{tab2}
\end{center}
\end{table}

\cref{fig:fanin_latency} shows the latency results for the fan-in experiments. On axis $x$, we have the number of executions during fan-in, as presented in \cref{fig:fanin}, while on axis $y$, we have the latency in seconds. As in previous experiments, every three measures framework displays horizontal lines, which means the performance does not decrease with the given executions. The slight variation in CWASI results in orange is due to resource usage already discussed in \cref{exp:fanout}. According to \cref{fig:fanin_latency} and \cref{tab2}, CWASI shows around 3 ms while OpenFaas performs around 8 ms and WasmEdge shortly over 110 ms. Overall, CWASI shows up to 60\% latency decrease compared to OpenFaas and up to 95\% when compared to WasmEdge.

In \cref{fig:fanin_throughput}, we see the throughput performance in the fan-in experiments. As expected from the fan-in latency results, the throughput shows three horizontal lines with slight variation in CWASI, in orange, due to the resource usage, where WasmEdge and OpenFaas remain constant throughout the experiment. On axis $x$, we have the executions; on axis $y$, we have the throughput in requests per second. CWASI displays around 300 requests per second while OpenFaas shows 130 and WasmEdge 8 requests per second. CWASI has 1.3x higher throughput than OpenFaas and over 30x higher throughput when compared to WasmEdge.

\section{Conclusion \& Future Work} \label{sec8}

We presented in this paper an interoperable inter-function communication model for the Serverless Edge-Cloud Continuum with three modes: Function Embedding, Local Buffer, and Networked Buffer. Additionally, we introduce CWASI, a WebAssembly runtime shim that implements the three-mode communication model to enable optimized inter-function communication. CWASI Shim leverages the function locality to identify and select the best inter-function communication mode. 

We evaluated CWASI by running Serverless workflows among the most relevant serverless invocation patterns sequence, fan-out, and fan-in using latency, throughput, and resource usage as the metric baselines. The experiments show that CWASI provides up to 95\% lower latency and up to 30x higher throughput compared to the start-of-the-art container runtimes for serverless platforms. CWASI shows significant improvement in serverless workflows that have high co-located inter-function communication.

Currently, CWASI only supports WasmEdge runtime. In the future, we plan to address this by extending CWASI for other runtimes to enable portability and flexibility. Additionally, CWASI relies on particular OCI specifications, i.e., annotations. To overcome this limitation, we plan to introduce programming models to identify functions of the same workflow and translate this composition into OCI specifications, such as arguments and annotations. We envision programming models identifying whether a serverless function is better suited for standard containers or WebAssembly and creating a specific deployment hint for each serverless function.
Furthermore, we envision CWASI as an enabler for a \textit{BaaSless} Serverless Computing where the framework will manage the backend-as-a-service infrastructure, further simplifying application development. Our future programming models will enable CWASI to identify necessary backend-as-a-service for a specific workflow. Once the BaaS' is identified, CWASI will proactively start and manage every BaaS necessary to run this particular workflow.

\begin{acks}
This research received funding from the EU’s Horizon Europe
Research and Innovation Program under Grant Agreement No.
101070186. EU website for TEADAL: \url{https://teadal.eu}.
\end{acks}

\balance

\bibliographystyle{unsrtnat}
\bibliography{intro}

\end{document}